\documentclass[prb,twocolumn,amsmath,amssymb,showpacs]{revtex4}
\usepackage{graphicx}

\begin{document}


\title{Origin of second-harmonic generation in the
incommensurate phase of $\rm \bf K_2SeO_4$}

\author{T.A. Aslanyan}
\email{aslanyan@freenet.am}
\affiliation{Institute for Physical
Research, Armenian National Academy of Sciences, Ashtarak-2,
378410 Armenia}



\received{18 December 2003}
\begin{abstract}
We show that a ferroelectric phase transition takes place in the
incommensurate phase of the $\bf\rm K_2SeO_4$ crystal. The
ferroelectric character of the IC phase explains the
second-harmonic generation observed in the corresponding
temperature range.
\end{abstract}
\maketitle The crystals of the $A_2BX_4$ family (about 20 crystals
)\cite{I} undergo two successive phase transitions on cooling: an
incommensurate (IC) transition at temperature $T_i$ with a
modulation vector near the $b/3$ point in the Brillouin zone, and
upon further cooling at temperature $T_c\,$ a commensurate lock-in
transition to the triple-period ferroelectric phase with the
modulation vector $b/3$. The most striking peculiarity of the
$\bf\rm K_2SeO_4$ crystal, which belongs to the $A_2BX_4$ family,
is the second-harmonic generation (SHG) in the IC phase,\cite{Yes}
of the same intensity magnitude as that in the $b/3$- phase. Low
temperature $b/3$- phases of the $A_2BX_4$ type crystals are
improper ferroelectrics, and therefore the SHG is a normal
manifestation of this feature. However, so far there was neither
 pointed out a plausible reason for the SHG in the IC
phase of $\bf\rm K_2SeO_4$, nor explained why it takes place not
in all the crystals of the $A_2BX_4$ family. It is worth noting
that the SHG was observed also in the IC phase of quartz by Dolino
and Bachheimer (1977)\cite{Dolino} and in the IC phase of the
ammonium fluoroberyllate $\bf\rm (NH_4)_2BeF_4$ by Alexandrov et
al. (1978)\cite{Alex}, and both observations also can be explained
using the theory developed below. The attempt to explain the SHG
in the IC phase of $\bf\rm (NH_4)_2BeF_4$ was carried out by
Golovko and Levanyuk\cite{Sec} based on the spatial dispersion of
the dielectric constant. However, the expected effect appeared to
be small, and besides, it is not explained why in such a case the
SHG is not observed in all the crystals with the IC phases.

In the present paper we show that within the IC phase of the
$\bf\rm K_2SeO_4$ crystal, necessarily  a transition to the
ferroelectric IC phase takes place. We analyze the exact solution
in frame of the Landau theory. The physical reason for such a
ferroelectric transition is the coupling between the crystal's
polar symmetry vibrational mode $P$ and some displacements of the
IC domain walls. We demonstrate that relative displacements of the
domain walls in the domainlike IC structure induce polarization
along the $z$ direction, and can be viewed as an additional phonon
branch of a polar symmetry. This mode is not explicitly introduced
in the equations below. It is a component of the IC phase $\varphi
(x)$ function, and is taken into account implicitly, in the
integral form. The coupling of the two modes renormalizes the
frequency of the lower mode down to zero within the IC range,
which induces a ferroelectric transition.

Structure of the IC modulation of $\bf\rm K_2SeO_4$ was defined in
the neutron diffraction by Iizumi \textit{et al}.\cite{Iizumi} The
thermodynamic potential describing the IC phase of $\bf\rm
K_2SeO_4$ was detailed studied by Sannikov and Levanyuk.\cite{San}
For the two-component order parameter with components
$\eta_1=\eta_0\cos u$ and $\eta_2=\eta_0\sin u$ it can be written
as:
\begin{eqnarray}
\nonumber &&\Phi  =\!\int dx\{\frac{\alpha}{2}\eta_0^2+
\frac{D}{2}\eta_0^2(\frac{\partial u} {\partial
x})^2+\sigma\eta_0^2\frac{\partial u} {\partial x}
+\frac{f}{2}\eta_0^6\cos 6u +\\ && B\eta_0^4+ F\eta_0^6+
 \frac{r}{2}P\eta_0^3\cos
3u +\frac{\chi^{-1}_0}{2}P^2 +hP^4-P\mathcal{E}\}
\end{eqnarray}
where $P$ is polarization vector in the $z$ direction,
$\mathcal{E}$ is electric field in the $z$ direction. In case of
$\sigma =0, \;D> 0$ this potential would describe a phase
transition from the symmetric high-temperature phase to the
triple-period commensurate phase at the temperature $\alpha =0$.
For $f<0$ the equilibrium values of the modulation phase $u$ are
$0,\, \pi/3, \,2\pi/3,\, \ldots$ (otherwise $\pi/6,\,
3\pi/6,\,\ldots$). However, in virtue of the Lifshitz invariant
($\sigma\neq 0$) the phase transition takes place from the
high-temperature phase to the IC phase at $\alpha >0$ with $u
=kx+\varphi$ and $k=-\sigma /D$.

Substituting $u =kx+\varphi (R)$ in potential (1) one obtains for
the $\varphi$- and $P$- dependent part of potential (1) the
following expression:
\begin{eqnarray} \nonumber
\tilde\Phi =\int dx\{ \frac{D}{2}\eta_0^2(\frac{\partial\varphi}
{\partial x})^2
+\frac{f}{2}\eta_0^6\cos (6kx+6\varphi )+\\
\frac{r}{2}P\eta_0^3\cos (3kx+ 3\varphi )
+\frac{\chi^{-1}_0}{2}P^2 +hP^4-P\mathcal{E}\}
\end{eqnarray}
Minimization of this expression with respect to $\varphi$ gives
the equation:
$$D\frac{\partial^2\varphi}{\partial x^2}+3f\eta_0^4\sin
(6kx+6\varphi)+\frac{3r}{2}P\eta_0\sin (3kx+ 3\varphi ) =0$$ After
single integration over $x$ and taking for the integration
constant the form $|f|\eta_0^4(1+c)$ one obtains:
\begin{eqnarray}
\frac{\partial u}{\partial x}=\sqrt{|f|/D}\,\eta_0^2\sqrt{1+c-\cos
6u +\frac{rP}{|f|\eta_0^3}\cos 3u},
\end{eqnarray}
 where $u=kx+\varphi$. The integration constant $c$, as well as the
polarization vector $P$, can be defined via minimization of the
potential with respect to these parameters after substituting the
solution of Eq. (3) in the potential (2).
\begin{figure}[t]
\mbox{\includegraphics[scale=0.9]{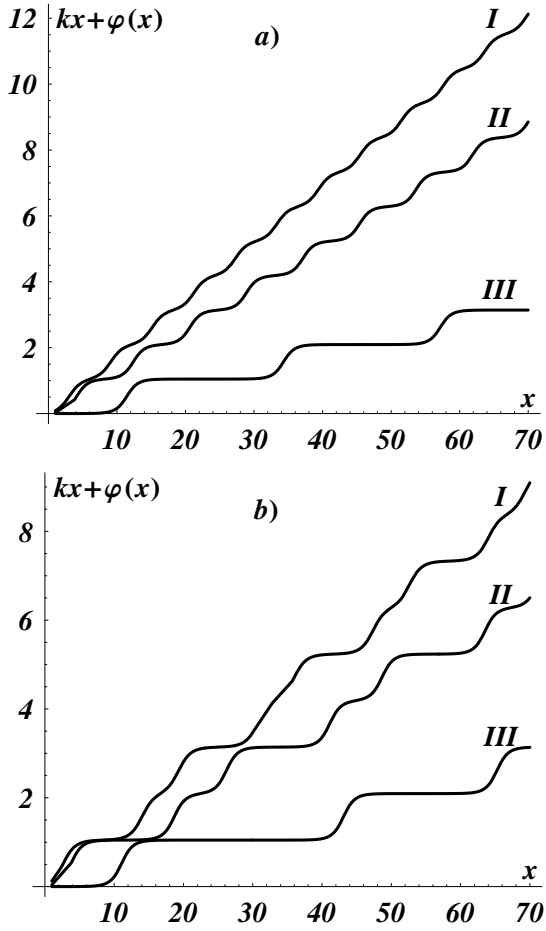}} \caption{a) Coordinate
dependence of the IC modulation phase for different temperatures.
The curves I , II, III represent various stages of evolving of the
domainlike modulation on cooling: curve I,
$\eta_0^2=0.21,\;c=0.138$; II, $\eta_0^2=0.23,\;c=0.010095$; and
III, $\eta_0^2=0.2326288,\;c=9\times 10^{-9}$ (the last is very
close in $\eta_0$ to the lock-in phase amplitude $\eta_0^\ast$).
The coordinate $x$ is given in multiples of the lattice parameter.
The height of each step on the curves is equal to $\pi/3$. Each
domain (plateau) has nonzero polarization vector in the $z$
direction. However, the neighboring plateaus are polarized in
opposite directions, providing crystal's zero integral
polarization. b) Transformation of the curves I , II, III induced
by the electric field, applied in the $z$ direction. The effect of
the electric field $\mathcal E$ comes to the increasing of the
plateaus' areas if the polarization vector is in the direction of
$\mathcal E$, and to the reduction of those otherwise. The height
of each step remains $\pi/3$. Such a transformation should take
place spontaneously, without an external field.}
\end{figure}
Prior to considering the ferroelectric transition in this system
let us discuss the solutions of Eq. (3) for the case of $P=0$. The
solution to this kind of equations for the IC structures was
analyzed in the pioneer work by Dzyaloshinski \cite{Dz} (it was
also detailed studied by Golovko\cite{Gol}), and for Eq. (3), it
has the form:
\begin{eqnarray}\varphi(x) +kx = \frac{1}{3}\;{\rm
am}(3\sqrt{|f|/D}\, \eta_0^2\sqrt c\,x\,,\;-\frac{2}{c}),
\end{eqnarray}
where $\,\rm am\,$ is the \textit{elliptic amplitude function} and
$k$ is the IC vector at the temperature of the IC transition. (In
the present paper the notations of the Wolfram Research,
"Mathematica" program for the elliptic functions  are used.)
Substituting solution (4) in the potential (2) and minimizing over
the parameter $c$ one obtains:
\begin{subequations}
\begin{eqnarray}
2\sqrt{|f|/D}\,\eta_0^2\sqrt c\,\hbox{\rm E}(-\frac{2}{c})= \pi
k\; ;
\end{eqnarray}
\mbox {$\rm E$} is the {\textit{complete elliptic integral
of the second kind},} which has the asymptotic\\[-10pt]
$$\sqrt c\,{\rm
E}(-\frac{2}{c})\sim \sqrt 2 (1 -\frac{c}{4}\ln c)\;\;{\rm
for}\;\;c\to 0.$$ Eq. (5a) can be numerically solved with respect
to $c$ for given coefficients $f,\,D,\,k,\,\eta_0$, and one
obtains that the parameter $c$ monotonously decreases from
$+\infty$ at $\eta_0=0$ (i.e., at the IC transition $T_i$) down to
$c=0$ at the continuous lock-in transition $T_c$. As it follows
from Eq. (5a), at the lock-in transition point the IC amplitude is
tending to
\begin{eqnarray}
\eta_0^2\to {{\eta_0^\ast}^2}=\frac{\sqrt 2 \pi
k}{4}\sqrt{D/|f|}\;\;{\rm with}\;\; c\to 0,
\end{eqnarray}
\end{subequations} which gives that $k\sim {\eta_0^\ast}^2$,
if the remaining coefficients are of the same order of magnitude.

In Fig. 1(a)  the solutions given by Eq. (4) for $k=2\pi /30$,
$D/|f|=1$ and $P=0$ for the three values of the IC amplitude are
plotted. At the IC transition $T_i$ ($c\to\infty$) the
corresponding graph must look like a straight line with slope $k$
($\varphi =const$, i.e., sinusoidal modulation), and on cooling it
acquires the domain-like features. Due to the symmetry, each point
$x$ on the curves of Fig. 1(a) is associated with a nonzero
polarization, which is spatially distributed as $P\sim\eta_0^3\cos
(3kx+ 3\varphi )$ with zero-integral effect. Meanwhile, each
domain there has a nonzero-integral polarization. In the areas of
any of two neighboring domains $\varphi +kx \approx n\pi/3$ and
$\varphi +kx \approx (n+1)\pi/3$ respectively, giving rise to the
$P$-s of opposite signs.

In Fig. 1(b) the transformation of the curves of Fig. 1(a) under
the applied electric field $\mathcal E$ is shown. The field
$\mathcal E$ is taken into account via replacing in Eq. (3) of the
polarization vector $P$ by $P=\chi_0\mathcal E$. An identical
transformation of the domains takes place in the presence of
nonzero polarization $P$ in crystal (when $\mathcal E =0$).
Corresponding solutions were found both analytically and
numerically (using the "Mathematica" software) with the same
result. Note that the integral polarization vector contributed by
the nonequal domains in Fig. 1(b) is not zero.

Now we show, that the above described transformation of the domain
structure from that shown in Fig. 1(a) to  Fig. 1(b) takes place
spontaneously, i.e., without applying an external electric field.
In other words, there takes place a ferroelectric phase transition
inside the IC phase. Let us calculate the dielectric
susceptibility of the crystal in the IC phase:
$$\chi =\frac{\partial P}{\partial \mathcal E}=\left
(\frac{\partial^2\tilde\Phi}
{\partial P^2}\right)_{P=0}^{-1}$$ Differentiation of potential
(2) gives:
\begin{equation}
\left (\frac{\partial^2\tilde\Phi}{\partial P^2}\right
)_{P=0}=\chi_0^{-1} - \frac{3r\eta_0^3}{2t}\int\limits_0^{t} \sin
3u\left (\frac{\partial u}{\partial P}\right )_{P=0}dx,
\end{equation}
$t=2\pi /q$ is the IC period, given by Eq. (10) below. In order to
calculate the derivative  $\left(\partial u /\partial
P\right)_{P=0}$, we introduce notation
\[{\mathbb
F}(u,\,P)=\int_0^u\frac{dv}{\sqrt{1+c-\cos 6v
+\frac{rP}{|f|\eta_0^3}\cos 3v}}\;.\] Integration of Eq. (3) gives
in this notation, ${\mathbb F}(u,\,P)-\sqrt{|f|/D}\,\eta_0^2x=0$,
from which follows that $\partial {\mathbb F}/{\partial P}+
\left({\partial{\mathbb F}}/{\partial u}\right){\partial
u}/{\partial P}=0$. The latter allows one to calculate the
derivative $\left({\partial u}/{\partial P}\right)_{P=0}$ via
calculation of the corresponding derivatives of the function
${\mathbb F}$ in the integral form (see the Appendix). After
substituting of the calculated derivative in Eq. (6), one obtains:
\begin{eqnarray}\nonumber
\left (\frac{\partial^2\tilde\Phi}{\partial P^2}\right
)_{P=0}=\chi^{-1}_0 -\frac{3\sqrt D \,
r^2}{4t|f|^{\frac{3}{2}}\eta_0^2c}\int\limits_{0}^{2\pi /3}\frac{
\sin^2 3u\,du}{\sqrt{1+c-\cos 6u}}=\\
\chi^{-1}_0 -\frac{\pi\sqrt D \,
r^2}{4t\eta_0^2|f|^{\frac{3}{2}}c^{\frac{3}{2}}}\;{_2{\rm F}_1
}\left(\frac{1}{2}\, ,\; \frac{3}{2}\, ,\; 2 ,\;
-\frac{2}{c}\right)\;\;
\end{eqnarray}
where ${_2{\rm F}_1 } \left({1}/{2}\, ,\; {3}/{2}\, ,\; 2 ,\;
-{2}/{c}\right)$ is a \textit{hypergeometric function}. This
hypergeometric function has the asymptotic
$${_2{\rm F}_1 }\left(\frac{1}{2}\, ,\; \frac{3}{2}\,
,\; 2 ,\; -\frac{2}{c}\right)\approx \frac{\sqrt {8c}}{\pi}\;\;
{\rm for}\;\; c\to 0 ,$$ which well fits the function in all the
range $0<c<1$, i.e., not only at $c\to 0$. As one can see from Eq.
(10), the IC period has an asymptotic: $c\to 0$, $\;t\sim |\ln
c|/\eta_0^2$ (as shown below, on approaching the lock-in
temperature $T_c$ the coefficient $c$ is tending to zero as $c|\ln
c| \sim |T-T_c|$).

So, for the susceptibility $\chi$ one obtains:
\begin{equation}
\chi^{-1}=\chi^{-1}_0 -\frac{\pi r^2\;{_2{\rm F}_1 }}{16|f|c\,{\rm
K}}\approx \chi^{-1}_0 - \frac{ r^2}{|f|\;c\,|\!\ln c|}
\end{equation}
and the stability coefficient $\chi^{-1}$ necessarily vanishes
(since the $c$ -dependent term in Eq. (8) is negative and
diverging with $c\to 0$), inducing ferroelectric phase transition
at the temperature $\chi^{-1}=0$. A loss of stability necessarily
takes place even for the case of large $\chi^{-1}_0$ (i.e., even
for a not soft polar mode of the high temperature phase), and it
is expected in the medium temperature range of the IC phase ($c
\lesssim 0.1$).

However,  some alternative situations are possible near the loss
of stability $\chi^{-1}=0$. As it follows from Eq. (3), the
polarization $P$ which appears below the transition should be
sufficiently small:
\begin{eqnarray}
P<c\,\eta_0^3|f|/r
\end{eqnarray}
Otherwise, the expression under the root in Eq. (3) takes negative
values for some $u$, what means that the crystal does not fall
within the IC phase, but within the commensurate lock-in phase
instead, skipping the ferroelectric IC phase. In terms of Fig.
1(b), it is equivalent to the infinite increasing of the one
domain's size, which covers all of the crystal area. The right
hand side of Eq. (9) is very small, since $\eta_0^3$ in Eq. (9) is
a small parameter, and $c$ tends to zero in the low-temperature
range of the IC phase (see also below). In the case of a
first-order ferroelectric phase transition giving rise to a large
$P$ (in Landau theory, it means a negative fourth-order term in
the potential's $P$- expansion), inequality (9) does not hold, and
the transition transfers the crystal directly to the lock-in
phase, skipping the ferroelectric IC phase. Even for some
order-disorder transitions, which are very close to the
second-order type (the polarization vector $P$ below the
order-disorder transitions is larger, due to the small Curie
constant compared to that for the displacement type) the expected
ferroelectric transition is more likely to turn into the
transition to the lock-in phase, skipping the ferroelectric IC
phase. Note that among the crystals of the $A_2BX_4$ family only
$\bf\rm K_2SeO_4$ is of a displacement type, and therefore for
this crystal the ferroelectric phase transition (from the IC to
the ferroelectric IC phase) can be considered as a most likely. In
other words, the IC structure of $\bf\rm K_2SeO_4$ becomes
ferroelectric with the modulation, spatially distributed like that
in Fig. 1(b). One should expect that the ferroelectric transition
takes place in the beginning of the domainlike modulation
formation, i.e., at temperatures near the curve I in Fig. 1(a),
when the $c$ coefficient falls below unity.

As it follows from Eqs.(5a, 5b), for estimates of inequality (9),
the asymptotic of the parameter $c\to 0$ can be used:
$$c\ln c \approx 8\frac{\eta_0^2- {\eta_0^\ast}^2}{\eta_0^2}
\approx 16\frac{a(T_c-T)}{\eta_0^\ast};$$ near the lock-in point
we presented $\eta_0- {\eta_0^\ast}\approx a(T_c-T)$, where $a$ is
some constant. So, Eq. (9) can be presented as
$P<16a(T-T_c)\,\eta_0^2|f|/r$, when $T\to T_c$ (we neglected a
logarithmically diverging term $\ln c$, which does not change
order of magnitude of the estimated values, since, as discussed
below, $|T_c-T|$ can not be sufficiently small for it).

For obtaining of the inharmonic fourth order term in the $P$
expansion of the potential one should calculate the derivative
$\left ({\partial^4\tilde\Phi}/{\partial P^4}\right )_{P=0}$. We
shall introduce here only the result of this calculation. In
addition to the $hP^4$ term in Eq. (2), one obtains a negative,
strongly diverging as $c^{-3}$ with $c\to 0$ term. If the sum of
the inharmonic $P^4$ terms becomes  negative earlier than
$\chi^{-1}=0$, then takes place a first-order transition. The
latter makes more probable the first order transition, even if the
coefficient $h>0$ is not small.

Now we shall discuss some features of the IC phase for $P=0$ case
in order to compare it with behavior of the ferroelectric IC
phase. To study the temperature evolving of the ferroelectric IC
structure upon cooling one should minimize potential (2) also over
the IC amplitude $\eta_0$. However, we imply in the present paper
only increasing of the IC amplitude $\eta_0$ upon cooling, and do
not discuss the character of the lock-in transition.

First we discuss some structure peculiarities of the
non-ferroelectric IC phase of $\bf\rm K_2SeO_4$. The length of
each plateau in Fig. 1(a) is six times shorter than the IC
wavelength $2\pi /q$, where $q$ is the IC wavevector [$q$ is equal
to the average slope of the given curve in Fig. 1(a)]. From this
geometrical fact it follows, that the IC modulation argument
$u=kx+\varphi$ can be presented as $u =qx+\theta (x)$, where
$\theta (x)$ is a periodic function with period $2\pi /6q$. As it
follows from Eq. (4), roughly $\theta (x)$ can be approximated by
$\sin 6qx$. The IC satellite reflections observed in the X-ray
diffraction are induced by the Fourier-components of the
modulation function $\eta_0\cos [(\frac{b}{3}+q)x+\theta (x)]$.
The IC satellites observed for such a modulation should appear in
the points $\frac{b}{3}+q$; $\frac{b}{3}+q \pm 6q,\;\pm 12q,\;
\ldots$ near the main Bragg- reflections $\textbf G$. In other
words, the manifestation of the domain-like IC structure in
$\bf\rm K_2SeO_4$ in diffraction should be only the simultaneous
observation of the IC satellites of about the same intensity
(asymmetrically disposed with respect to the $\frac{b}{3}$ point)
in the points $\frac{b}{3}-5q$ and $\frac{b}{3}+7q$ near
$\textbf{G}$. Intensive IC satellites in the points $\frac{b}{3}+q
\pm 2q,\;\pm 3q,\;\;\pm 4q,\;\pm 5q$, which are observed in
diffraction, are not contributed by the domain features of the
modulation, and can not be viewed as a manifestation of a
domainlike IC structure. The satellites $\frac{b}{3}-5q$ and
$\frac{b}{3}+7q$ were not observed for $\bf\rm K_2SeO_4$, which
proves that the domainlike structure is not well developed in
$\bf\rm K_2SeO_4$, and the ferroelectric phase exists only in the
medium range of the IC phase. So, as it follows from the exact
solution of Eq. (3), any crystal of the $A_2BX_4$ type with well
developed IC domain structure (i.e., sufficiently small $c$)
should be ferroelectric. The ferroelectric transition does not
change the translation symmetry of the IC phase, and therefore the
satellites disposition for the ferroelectric IC phase should not
essentially differ from that for the $P=0$ case.

We also discuss the character of increasing of the IC modulation
period on cooling, which is logarithmically diverging, as it
follows from Eq. (4). The IC period for the modulation given by
Eq. (4) is
\begin{eqnarray}
t= \frac{2\pi}{q}=\frac{4\sqrt{D}\,}{\sqrt{|f|}\,\eta_0^2\sqrt
c}\;{\rm K}(-\frac{2}{c}),
\end{eqnarray}
where ${\rm K}(-{2}/{c})$ is the \textit{complete elliptic
integral of the first kind}. With $c\to 0$ the function ${\rm
K}(-{2}/{c})/\sqrt c$ has the asymptotic $$4{\rm
K}(-\frac{2}{c})/\sqrt c\sim \sqrt {2}\,(5\ln 2 -\ln c),$$ and $$
t\sim \frac{\sqrt{D}|\ln c|}{\sqrt{|f|}\eta_0^2}\sim \frac{|\ln
(T-T_c)|}{{\eta_0^\ast}^2}\sim \frac{|\ln (T-T_c)|}{k}$$ which is
logarithmically diverging. The temperature $T$ can approach the
lock-in temperature $T_c$ not closer than the temperature
fluctuation $\sqrt{<(\Delta T)^2>}$ , i.e.,
$$|T-T_c|>\;\sqrt{<(\Delta T)^2>}\;{\rm and}\; <(\Delta T)^2>=
\frac{T^2}{V c_v},$$ where $V$ is the crystal volume and $c_v$ is
the heat capacity. So, for any crystal of a finite size ($V\sim
10^{22}a^3$), the period $t$ can increase only about 10 times
compared to $2\pi /k$, in virtue of the logarithmical character of
the divergence (it is also confirmed by our numerical
simulations). Besides, in the real experiment the temperature $T$
can not approach the lock-in temperature $T_c$ even closer than
the instrumental temperature resolution $\Delta T \sim
10^{-3,\,-4}K$, which does not allow to observe any diverging IC
period, and the lock-in transition always should be observed as a
discontinuous phase transition with jump in the wavevector $q$
down to $q=0$.

For calculation of the ferroelectric IC modulation's period [the
$x$-length of the six neighboring domains in Fig. 1(b)], one
should integrate Eq. (3) from $u=0$ through $u=2\pi$, which gives:
\begin{equation}
t=3\sqrt {\frac{D}{|f|}}\,\int\limits_{0}^{\frac{2\pi}{3}}\frac{
du} {\eta_0^2\sqrt {1+c-\cos 6u+\frac{rP}{|f|\eta_0^3}\cos
3u}}\approx\;
\end{equation}

\begin{eqnarray}
\frac{-i8\sqrt D}{\eta_0^2\sqrt{|f|c\,\Delta c}}\left[{\rm
F}\!\left( \frac{i\sqrt{ c}}{2}\,,\,\frac{32}{c\,\Delta c}
\right)\! + {\rm F}\!\left( \frac{i\sqrt{\Delta c}}{\sqrt
8}\,,\,\frac{32}{c\,\Delta c} \right)\!  \right]\;\,
\end{eqnarray}
where $\rm F$ is the \textit{elliptic integral}, and $\Delta c=
c-rP/(|f|\eta_0^3)$, i.e., the condition $\Delta c>0$ coincides
with Eq. (9), and therefore it always holds in the IC phase. The
parameter $\Delta c$ decreases down to zero with increasing of the
polarization vector $P$ on cooling. Integral (11) was calculated
exactly, and after being simplified to the form of Eq. (12) for
$\Delta c\ll c<1$. The first term in Eq. (12), which gives the
size of the larger domains in Fig. 1(b), is diverging with $\Delta
c\to 0$ as $\sim |\ln\Delta c|$. Meanwhile, the size of the
smaller domains is given by the second term in Eq. (12), and it is
always finite, though it also increases with cooling.

Summarizing, we showed that in the IC phase of the $\bf\rm
K_2SeO_4$ crystal  takes place ferroelectric phase transition,
which explains the SHG in the IC phase of this crystal. Formation
of the ferroelectric IC phase is most likely near the
displacement-type phase transitions, though it can occur also in
those of the order-disorder type. The IC period increases on
cooling towards the lock-in temperature $T_c$ as $|\ln (T-T_c)|$,
though the size of the domains, polarized opposite to the
prevailing polarization, remains finite.
\appendix*{}
\section{Calculation of $\left({\partial u}/{\partial
P}\right)_{P=0}$}

In order to calculate the derivative $\left({\partial u}/{\partial
P}\right)_{P=0}$ one should calculate the corresponding
derivatives of the function ${\mathbb F}(u,\,P)$. So one obtains
$$\left(\frac{\partial {\mathbb F}}{\partial
u}\right)_{P=0}=\frac{1}{\sqrt{1+c-\cos 6u}}$$ and
$$\left(\frac{\partial {\mathbb F}}{\partial P}\right)_{P=0}=
-\frac{r}{2|f|\eta_0^3}\int\limits_{0}^{u}\frac{\cos 3v\,
dv}{(1+c-\cos 6v )^{3/2}}=$$
$$-\frac{r}{6c|f|\eta_0^3}\frac{\sin 3u}{\sqrt{1+c-\cos 6u}}.$$
We took into account that for the equilibrium value of $c$, the
condition $c(P)=c(-P)$ holds, from which follows that
$\left({\partial c}/{\partial P}\right)_{P=0}=0$. Similarly
$\left({\partial \eta_0 }/{\partial P}\right)_{P=0}$ is also equal
to zero. So, one obtains:
$$\left(\frac{\partial u}{\partial
P}\right)_{P=0}=\! -\left(\frac{\partial {\mathbb F}}{\partial
P}\right)_{P=0}\!\mbox{\Large/}\left(\frac{\partial {\mathbb
F}}{\partial u}\right)_{P=0} = \frac{r}{6c|f|\eta_0^3}\sin 3u$$
For integration over $x$ in Eq. (6), the following relations are
also used:
$$\int\limits_0^{t}dx= \int_0^{2\pi}\left(\frac{\partial
u}{\partial x}\right)^{-1}{du}=\frac{3\sqrt D}{\sqrt{ |f|}\,
\eta_0^2}\int\limits_0^{2\pi /3}\frac{du}{\sqrt{1+c-\cos 6u}},$$
where we used the right hand side of Eq. (3) for $\left({\partial
u}/{\partial x}\right)$ at $P=0$.


\begin{thebibliography}{widest-label}
\bibitem{I}
Y. Ishibashi,  {\it Incommensurate Phases in Dielectrics}, edited
by R. Blinc and A.P. Levanyuk, Modern Problems in Condensed Matter
Sciences, Vol.{\bf 14.2}, North-Holland (1986).
\bibitem{Yes} A.M. Arutyunyan, B. Brezina, S.Kh. Esayan, V.V.
Lemanov, Sov. Phys., Solid State, \textbf{24}, 814, (1982).
\bibitem{Dolino}
G. Dolino, J.P. Bachheimer, phys.stat.sol. (a), \textbf{41}, 673,
(1977).
\bibitem{Alex}
K.S. Alexandrov, A.N. Vtyurin, V.F. Shabanov, Sov. Phys., Pis'ma
JETP, \textbf{28}, 153, (1978).
\bibitem{Sec}
V.A. Golovko, A.P. Levanyuk, Sov. Phys., JETP, \textbf{50}, 780,
(1979).
\bibitem{Iizumi}
M. Iizumi, J.D. Axe, G. Shirane and K. Shimaoka, Phys. Rev.
\textbf{B 15}, 4392 (1977).
\bibitem{San}
D.G. Sannikov, A.P. Levanyuk, Sov. Phys., Solid State, \textbf{4},
1005, (1978).
\bibitem{Dz}
I.E. Dzyaloshinski, Sov. Phys., JETP, \textbf{47}, 336, 992,
(1964).
\bibitem{Gol}
V.A. Golovko, Sov. Phys., Solid State, \textbf{22}, 2960, (1980).
\end{thebibliography}
\end{document}